\documentclass[runningheads]{llncs}
\usepackage{graphicx}
\newcommand{\addition}[1]{{\color{black} #1}}
\newcommand{\modification}[1]{{\color{black} #1}}

\usepackage{subfigure}
\usepackage{cite}
\usepackage{xcolor}
\usepackage{url}

\usepackage[utf8]{inputenc}

\begin{document}

\title{MBWU: Benefit Quantification for Data Access Function Offloading}
\author{Jianshen Liu\inst{1} \and
Philip Kufeldt\inst{2} \and
Carlos Maltzahn\inst{1}}
\authorrunning{J. Liu et al.}
\institute{University of California, Santa Cruz, Santa Cruz CA 95064, USA \and
Seagate, Longmont CO 80303, USA\\
\email{\{jliu120,carlosm\}@ucsc.edu}\\
\email{philip.kufeldt@seagate.com}}

\maketitle

\modification{

\begin{abstract}

The storage industry is considering new kinds of storage devices that support data access function offloading, i.e. the ability to perform data access functions on the storage device itself as opposed to performing it on a separate compute system to which the storage device is connected. But what is the benefit of offloading to a storage device that is controlled by an embedded platform, very different from a host platform? To quantify the benefit, we need a measurement methodology that enables apple-to-apple comparisons between different platforms. We propose a Media-based Work Unit (MBWU, pronounced "MibeeWu"), and an MBWU-based measurement methodology to standardize the platform efficiency evaluation so as to quantify the benefit of offloading. To demonstrate the merit of this methodology, we implemented a prototype to automate quantifying the benefit of offloading the key-value data access function.

\end{abstract}

\section{Introduction}

Benefit quantification is critical in value assessment of offloading data access functions from traditional host platforms to embedded platforms that are expected to serve beyond the role of transitional storage devices. Though a couple of frameworks focusing on breaking down the offloading complexity ~\cite{gu2016biscuit, phothilimthana2018floem} have been proposed in recent research, the fundamental question regarding how much can be saved from offloading a given data access function to an embedded platform has not been addressed. The challenge is whether to offload a data access function depends not only on the characteristics of workloads, which essentially are the function calls organized in some pattern, but also on the performance of the storage media with which the data access function interacts. In practical environments, hardware platforms and workloads of interest may differ significantly; solutions of benefit quantification focusing on specific functions ~\cite{wang2016ssd, kang2013enabling} or using simplified evaluation models ~\cite{boboila2012active, kim2011fast} may not apply to a different function. Furthermore, since different storage media have significantly different requirements on the bandwidth of various platform resources, the optimal placement of data access functions in terms of the platform efficiency can be dramatically different. We propose a Media-Based Work Unit (MBWU, pronounced "MibeeWu") and developed an MBWU-based measurement methodology for the purpose of standardizing the efficiency comparison for different platforms running a given workload over a specific storage media. By evaluating the efficiency of each platform in terms of its cost (\$/MBWU), power (kW/MBWU), and space ($m^{3}$/MBWU), we can quantify the benefit of offloading a data access function from traditional host platforms to embedded platforms. We have implemented a prototype for evaluating key-value data management function offloading and generated instructive results from our experiment. We discuss MBWU as well as this measurement methodology in detail in section \ref{Methodology}.

Starting from Active Disks ~\cite{riedel1998active, keeton1998case, riedel1997active, tiwari2013active, ouyang2013active}, moving high-selectivity data access functions to storage devices gains increasing research interest mainly because of the conceivable benefits ~\cite{kang2019towards} such as reducing the size of data transmission between hosts and storage devices, reducing total power consumption, increasing overall resource utilization, and simplifying the application design by leveraging high-level storage semantics. For example, key-value smart storage devices can substitute the translation layers from key-value down to physical block address, which includes a key-value to file translation in the front, a file to logical block address translation in the middle, and a logical block address to physical block address translation at the bottom. Besides these benefits, energy consumption is thought to be another major saving from offloading functions to storage devices. For example, Choi et al. ~\cite{choi2015energy} identified more than 3x energy efficiency with 80\% compute offloaded for data-intensive applications.

Though various benefits have been studied, the storage industry remains conservative when adding data access functions to storage devices. The main barrier is that extra processing required in the storage device increases the cost of the device. Since applications run on system platforms, we believe an increase in storage device does not necessarily increase the overall platform cost. Considering the variety of workloads and the diversity of hardware, we need a systematic and reproducible methodology to quantify the overall benefit of offloading any given data access function to embedded platforms.

\section{The MBWU-based Methodology} \label{Methodology}

\subsection{Background}

The emergence of various storage technologies has changed the regular formula for constructing storage infrastructure. Historically, this formula was built around hiding the latency of storage devices using caching. However, innovations of recent NAND and storage-class memory technologies (e.g., V-NAND ~\cite{kang2016256}, 3D XPoint ~\cite{wiki:3dxpoint}) have altered the cost-optimal placement of various software and hardware resources ~\cite{theis2017end, shulaker2017three} since storage media of different performance impose different demands for the bandwidth of CPU, memory, network, and storage interface. For example, applications may want memory closer to computation for slow storage media because hiding data access latency is important, while the applications may want storage closer to computation for fast media because high-speed networking fabrics and data buses are expensive. With more domain-specific processing units (e.g., GPU, Google TPU ~\cite{wiki:tpu}, FPGA) taking over computations from host CPUs, the storage industry asks itself the same question: What should be done to improve the cost-efficiency of utilizing a specific storage media for data access? In terms of the placement of data access functions, the specific question is: For a given workload, can offloading a data access function from host platforms to storage devices reduce the overall cost per performance when the workload uses the same storage media?

\subsection{What is MBWU}

Host platforms and embedded platforms differ significantly in cost, usage, performance, power, and form factor. To compare the cost per performance of different platforms, we need to have a reference point to normalize the performance value generated from heterogeneous platforms so that these normalized values are directly comparable. The reference point is required to be \textbf{platform-independent} but \textbf{media- and workload-dependent}. The reasons are as follows:

\begin{itemize}
\item \textbf{platform-independence}: The reference point should be platform-related hardware independent. Otherwise, the normalized performance value of a platform may not be able to represent the efficiency of the platform utilizing a specific storage media under a workload. For example, if the reference point relates to a specific CPU architecture, then the normalization for the performance of a platform using a different CPU is skewed by the efficiency difference caused by the different CPUs.

\item \textbf{media-dependence}: Since the cost-optimal placement of functions is sensitive to types of storage media, the reference point should be media-dependent so that we can always normalize the performance of different platforms to the efficiency of utilizing a specific storage media. For this to work, all the different platforms under test should use the same type of storage media for performance evaluation.

\item \textbf{workload-dependence}: From an application point of view, the performance of a platform is the amount of work the platform can do in a unit of time. To measure the amount of work done, we need to define a unit of work as the reference point so that the performance of different platforms can be normalized to the number of units they can perform. Since different workloads have different work definitions, the unit of work should be measured in workload operations (WOs). Hence, the reference point is workload-dependent.
\end{itemize}

\noindent
The combination of platform-independence and media-dependence indicate that the reference point can only be media-based. We call this media-based and workload-dependent reference point \textbf{MBWU} and define it as the highest number of workload operations per second (WOPS) a given workload on a specific storage media can achieve with all external caching effects eliminated/disabled. In this definition, workload operations should not be simply interpreted as block I/Os. For key-value operations as the workload, a WO is a \texttt{GET}, \texttt{PUT}, or \texttt{DELETE}. For file operations as the workload, a WO is a \texttt{read} or \texttt{write}. On the other hand, we use the term storage media to express a configuration of storage devices. For example, a storage media can be a device with six flash chips, or a device that combines a spinning media, two flash components, and some non-volatile random-access memory. The MBWU definition has no requirement on what the storage media should look like, which means our MBWU-based measurement methodology is seamlessly applicable to different types of storage media. In the following sections we use the two terms \textit{storage media} and \textit{storage device} interchangeably. Finally, since an MBWU only depends on a specific storage media and a given workload, its WOPS should only be throttled by a specific storage media. Platform-related system resources like CPU, memory, and network can throttle the WOPS when measuring an MBWU. Resources like memory can also enhance the WOPS when the data access is from memory instead of storage devices. A throttled or enhanced WOPS number is not an MBWU because it is platform-dependent.

Once an MBWU is measured, the performance of a platform can be measured by its maximum MBWUs under the same workload. The greater number of MBWUs a platform can generate, the higher performance the platform is. The efficiency of a platform can then be evaluated based on the cost, power, and space of this platform. For example, if the platform can generate \textit{M} MBWUs, cost-efficiency of the platform (\$/MBWU) can be calculated by $ \frac{cost(plat)}{M} $. Similarly, the power-efficiency of the platform (kW/MBWU) can be calculated by $ \frac{amp \cdot volt}{1000 \cdot M} $, where \textit{amp} and \textit{volt} are the current and voltage of the platform respectively when it was generating \textit{M} MBWUs. Space-efficiency of the platform ($m^{3}$/MBWU) can be calculated in a corresponding manner by $ \frac{volume(plat)}{M} $.

The MBWU-based measurement methodology is intended to be used by storage device and storage system designers to assess whether to pair a given function to a specific storage media. It is not intended to be used to evaluate online methods during production workloads, because ensuring the workloads are the same for different platforms is difficult.

\subsection{How to Measure MBWU(s)}

Measuring an MBWU for a given workload that uses a specific storage media is different from measuring MBWUs for a platform except that the storage media and workload should be the same in these two measurements. To measure a single MBWU, we need a capable host to drive the peak performance of a given workload running on a single storage device with all external caching effects disabled. Since high-selectivity workloads are primarily I/O bound, looking for such a host is not difficult. The process of measuring MBWUs for a platform is to measure the maximum steady-state WOPS of the same workload running on the platform with normal caching configuration. The goal of this measurement is to evaluate what is the maximum WOPS that is eventually limited by the platform-related resources instead of storage devices. One way to push the WOPS to the limit is to replicate the workload on multiple storage devices. Once we have the value of the maximum WOPS, the MBWUs of this platform is equal to this value divided by a single MBWU. Figure \ref{fig:MBWUs_to_devices} is an example to show the general relationship between the number of MBWUs and the number of storage devices. The increment of MBWUs brought by an additional storage device decreases as the device number increases until finally no increment exists on the total MBWUs. The final stable MBWUs is the MBWUs of this platform for the workload. One reason for the diminishing MBWUs increment shown in the figure is the increasing average CPU cycles on a single data read due to the increasing system memory pressure.

\begin{figure}[h!]
  \centerline{\includegraphics[scale=0.3]{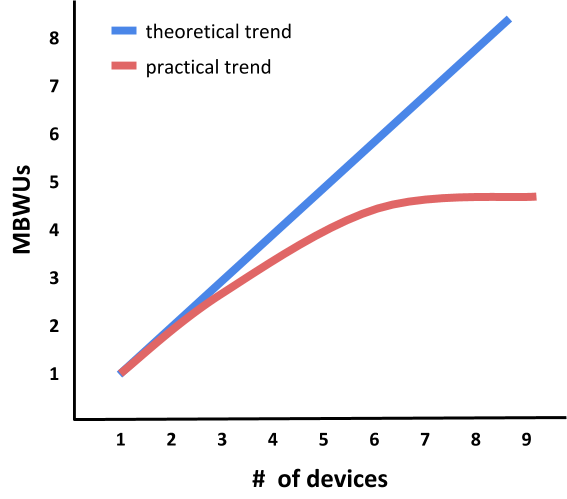}}
  \caption{Relationship between the Number of MBWUs and the Number of Storage Devices}
  \label{fig:MBWUs_to_devices}
\end{figure}

In addition to resources like CPU, memory, and network, the real estate issue is another important platform-related bottleneck. For example, a limitation on the available hardware connectors may limit the number of storage devices that can be attached to a platform, thus throttling the MBWUs of a platform as well. We have seen this type of limitation in our experiment (Section \ref{Evaluation}).

}

\subsection{Evaluation Prototype}

\begin{figure}[h!]
  \centerline{\includegraphics[scale=0.18]{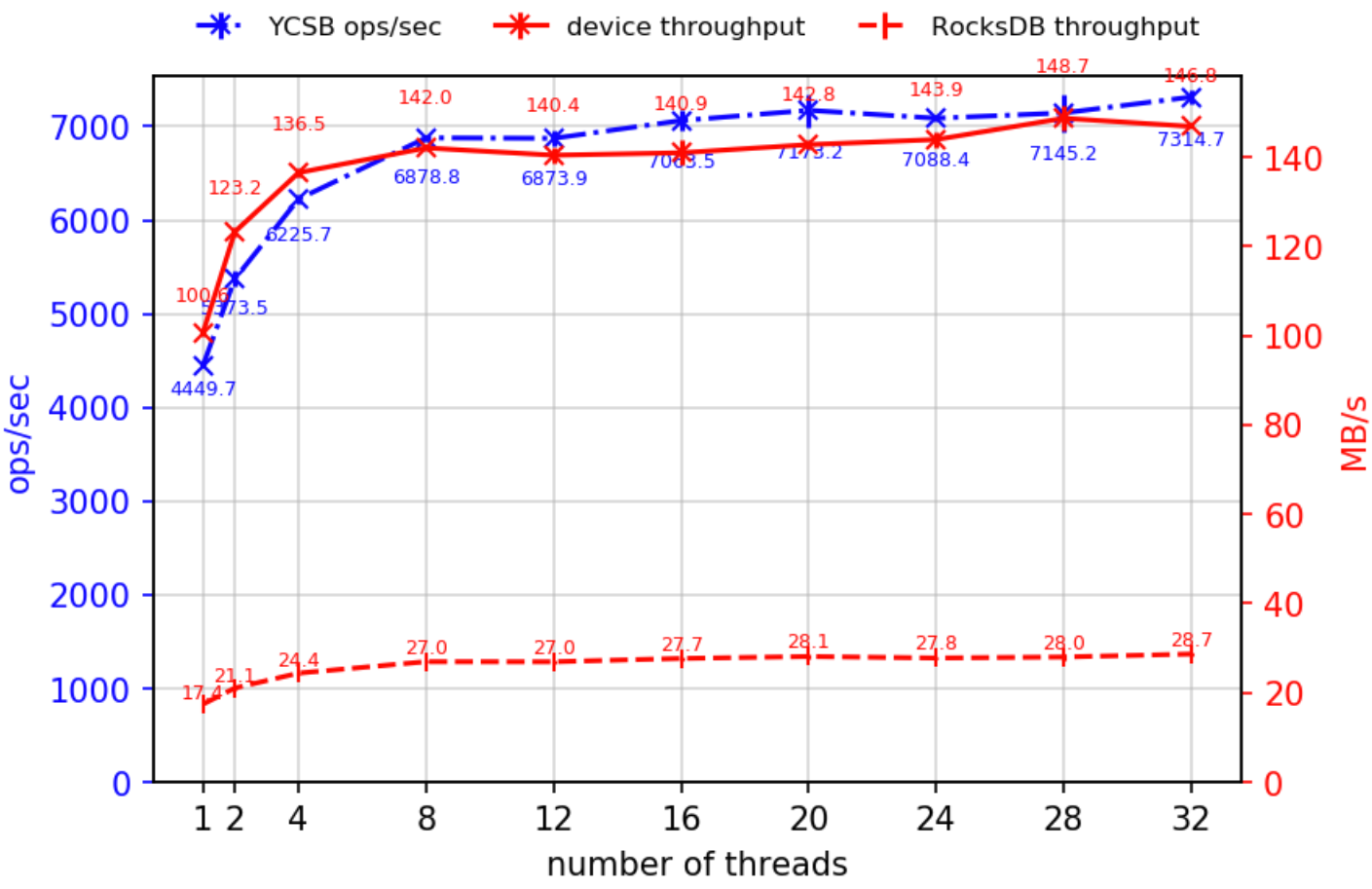}}
  \caption{Amplification of Key-value Data Traffic}
  \label{fig:data_traffic_amp}
\end{figure}

We chose key-value data management as a function to be offloaded in our study. The design goal of the evaluation prototype was to provide a framework to demonstrate the merit of our MBWU-based measurement methodology by automatically generate reproducible values that represent the benefit of offloading the key-value data management function for a given workload. Key-value data management is a typical high-selectivity function due to the massive amount of data needed to move back and forth between different levels of data representation in response to various operations in data management. For example, we used RocksDB ~\cite{borthakur2013under} as the key-value engines to run YCSB ~\cite{cooper2010benchmarking} workload A. We saw up to 6x traffic amplification between the key-value data received by the RocksDB (red dashed line) and the final data written out to the underlying block devices (red solid line) (Figure \ref{fig:data_traffic_amp}). Though we used the key-value function as an example, there is nothing to prevent the MBWU-based measurement methodology from being applied to other functions, such as read/write functions in the file system and SELECT/PROJECT functions in the database management system.

\begin{figure}[h!]
  \centerline{\includegraphics[scale=0.25]{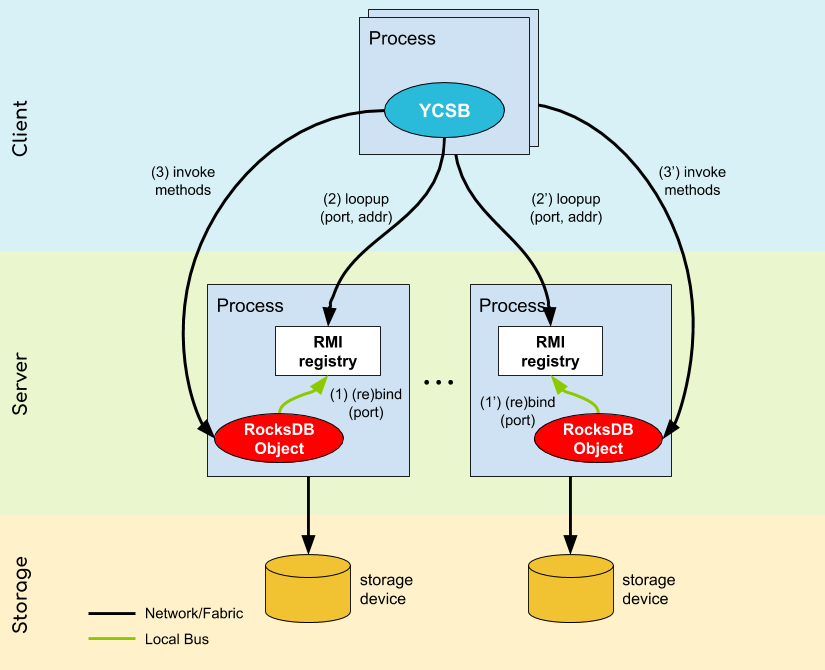}}
  \caption{Call Graph of RocksDB RMI Server}
  \label{fig:call_graph_rocksdb_rmi}
\end{figure}

\begin{figure}[h!]
  \centerline{\includegraphics[scale=0.4]{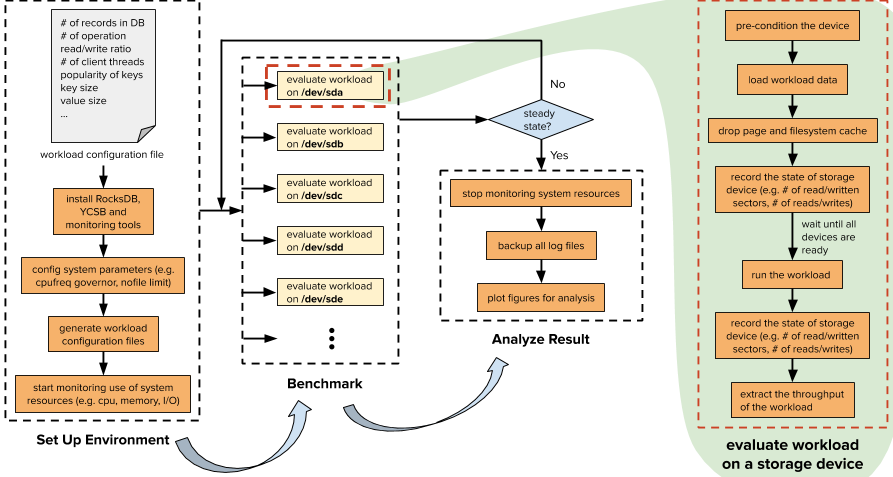}}
  \caption{A High-level View of the Evaluation Process}
  \label{fig:eval_process_hview}
\end{figure}

Our prototype starts with pre-conditioning for NAND-based storage devices used to store workload data. This process is necessary as it relates to whether reproducible results are possible. In the prototype, the pre-conditioning is implemented following SNIA performance test specification ~\cite{snia:pts18}; it purges the devices followed by performing a workload independent pre-conditioning process. After the storage devices are pre-conditioned, a number of RocksDB daemons are started and waiting for connection requests from YCSB. Storage devices, RocksDB daemons, and the YCSB processes are in a one-to-one relationship. Therefore, the number of RocksDB daemons is identical to the number of YCSB processes. The RocksDB daemon is implemented using Java RMI technology~\cite{wiki:javarmi}. It exposes all public interfaces (e.g., \textit{open()}, \textit{close()}, \textit{get()}, \textit{put()}, \textit{delete()}) of a RocksDB object to network securely by binding this object to an RMI registry (Figure \ref{fig:call_graph_rocksdb_rmi}). We have ported the RocksDB daemon program to support not only x86 and x86\_64, but also aarch64 platforms since most embedded platforms use ARM-based processors. A YCSB process looks up the corresponding RocksDB object from a specified RMI registry and requests to create a RocksDB instance on the host of the registry by issuing an \textit{open()} remote call. This call gives the RocksDB object the location of a RocksDB options file, which defines the shape of the internal LSM ~\cite{dong2017optimizing} tree and all the data migration policies for key-value data management. Having a consistent RocksDB options file for different platforms avoids using a "platform optimized" configuration file generated by RocksDB by default. Once RocksDB instances are successfully created, YCSB can start filling instances with initial key-value records to support later read/write operations. The data operation requests generated by YCSB are simply passed down by calling the exposed RocksDB interfaces. To ensure the final underlying LSM trees are consistent on platforms of different performance, we added an option to our prototype to control the speed of data loading. Slowing down the loads gives RocksDB instances enough system resources to finish regular data compaction and compression for keeping LSM trees stable. Finally, when the initial data are loaded, YCSB starts to run the workload specified by a parameter file with which the target workload is defined. YCSB offers various options to customize a workload: from the total number of operations, to request distribution, to the ratio between reads and writes, and so on. A high-level evaluation process of our prototype is shown in Figure \ref{fig:eval_process_hview}.

Depending on the configuration of a platform, the storage devices that RocksDB instances see can either be physical storage devices at local or network storage devices managed by any storage disaggregation protocol such as iSCSI ~\cite{wiki:iSCSI} and NVMe-oF ~\cite{minturn2015nvm}. The purge process, however, will always take place on the physical storage devices. We discuss different storage configurations and how they affect the cost-efficiency of a platform in Section \ref{Evaluation}.

Measuring MBWUs requires identifying which system resource is the bottleneck. Our prototype will automatically monitor and record utilization of CPU, memory, device I/O, network, and power for platforms during the whole evaluation process. At the end of a measurement, the prototype extracts all useful information from these logs and generates a platform resource utilization report for the target workload.

\section{Evaluation} \label{Evaluation}

The purpose of the evaluation is to demonstrate the use of our evaluation prototype discussed above for quantifying the benefit of offloading the key-value data management function from traditional host platforms to embedded smart key-value platforms. A smart key-value platform exposes a key-value interface instead of a block interface.

\subsection{Infrastructure}

\begin{figure}[h!]
  \centerline{\includegraphics[scale=0.18]{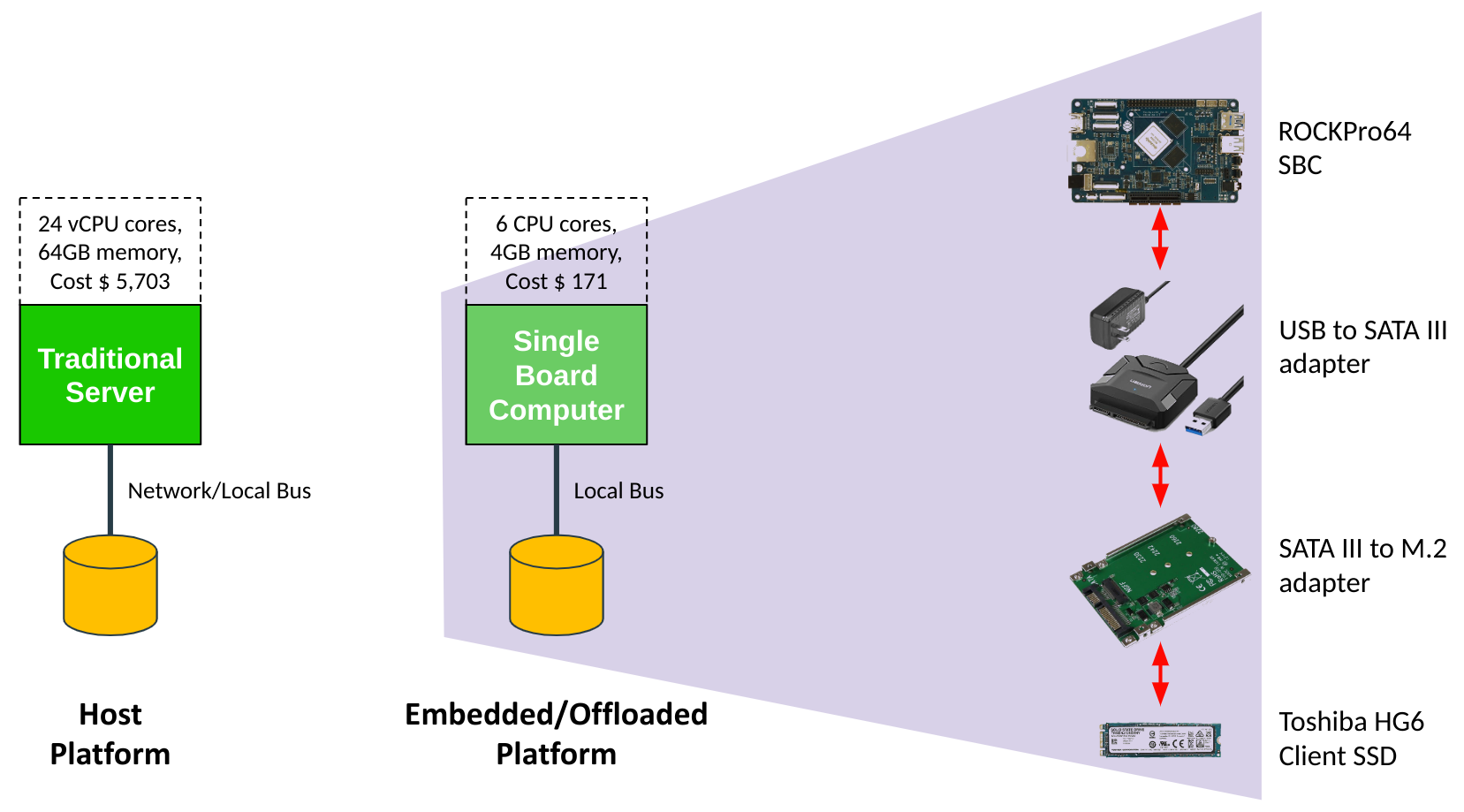}}
  \caption{Configuration of Our Host Platform and Embedded Platform}
  \label{fig:infra_setup}
\end{figure}

Figure \ref{fig:infra_setup} shows the basic components of the two platforms we set up for comparison. For the traditional platform, RocksDB runs on the host and stores data on either direct-attached storage devices or network storage devices. For the embedded platform, RocksDB runs on a single board computer (SBC) named ROCKPro64 ~\cite{rockpro6419}. This SBC, together with two adapters and a block storage device, creates a smart key-value storage device that clients can interact with through a key-value interface.

\subsection{Test Setup and Results}

\begin{figure}[h!]
  \centerline{\includegraphics[scale=0.17]{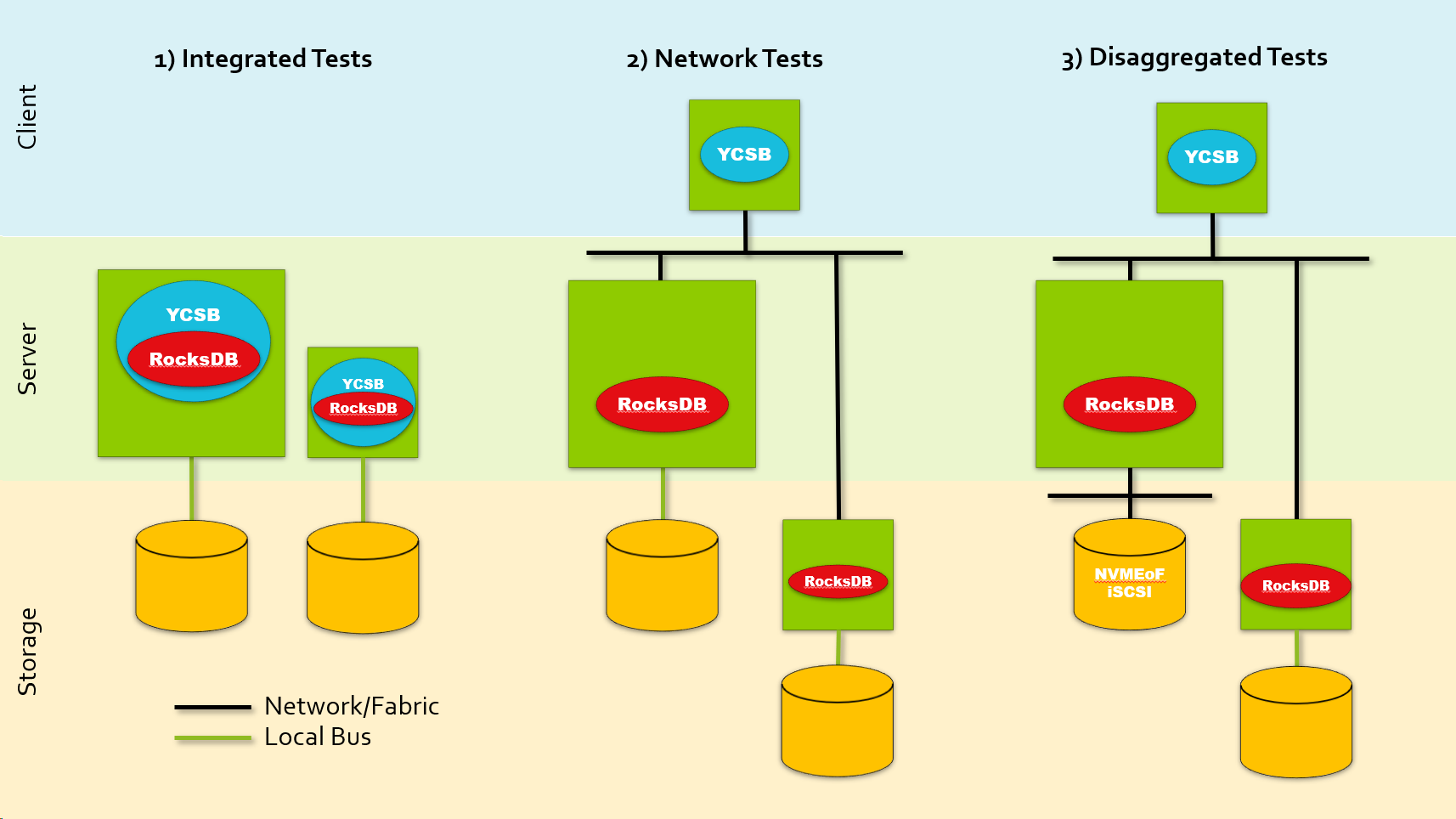}}
  \caption{Three-stage Test Setup}
  \label{fig:test_setup}
\end{figure}

To better understand where the benefit of offloading come from and how much savings occur regarding benefit, we have designed a set of tests involving three-stages (Figure \ref{fig:test_setup}). Each test was a different setup with different placement of either software or hardware components. We used the following workload in all tests. The key size is 16 bytes, and the value size is 4 KiB. The read/write ratio is 50/50 following a Zipf ~\cite{wiki:zipf} distribution for data accessing. Finally, the total size of dataset is 40 GiB.

\begin{figure}[h!]
  \centering
  \subfigure[Aggregated Throughput]{%
    \includegraphics[width=0.49\textwidth]{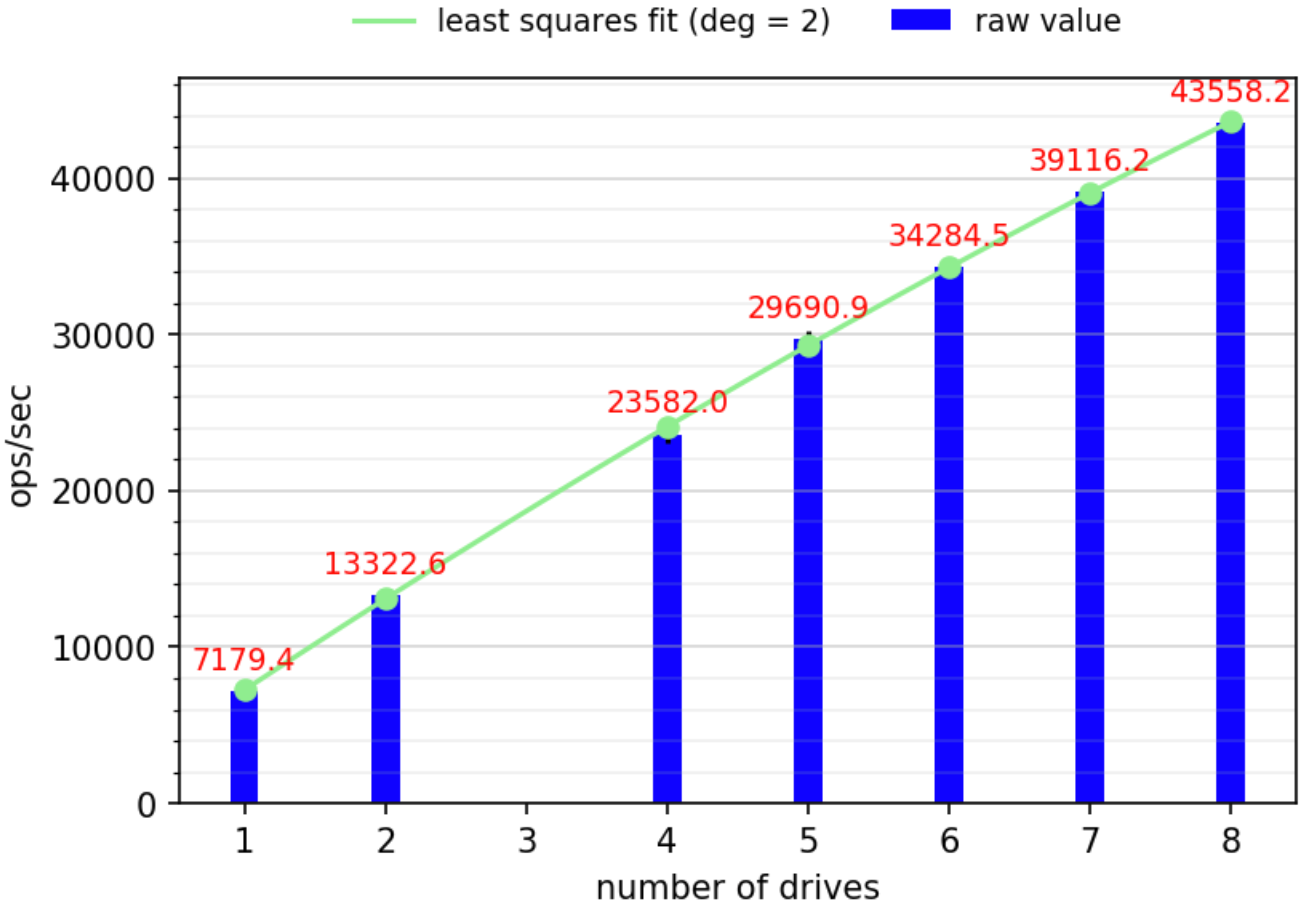}%
    \label{fig:integrated_test_throughput}%
    }\hspace{0.2cm}
    \subfigure[Platform Power Consumption]{%
    \includegraphics[width=0.49\textwidth]{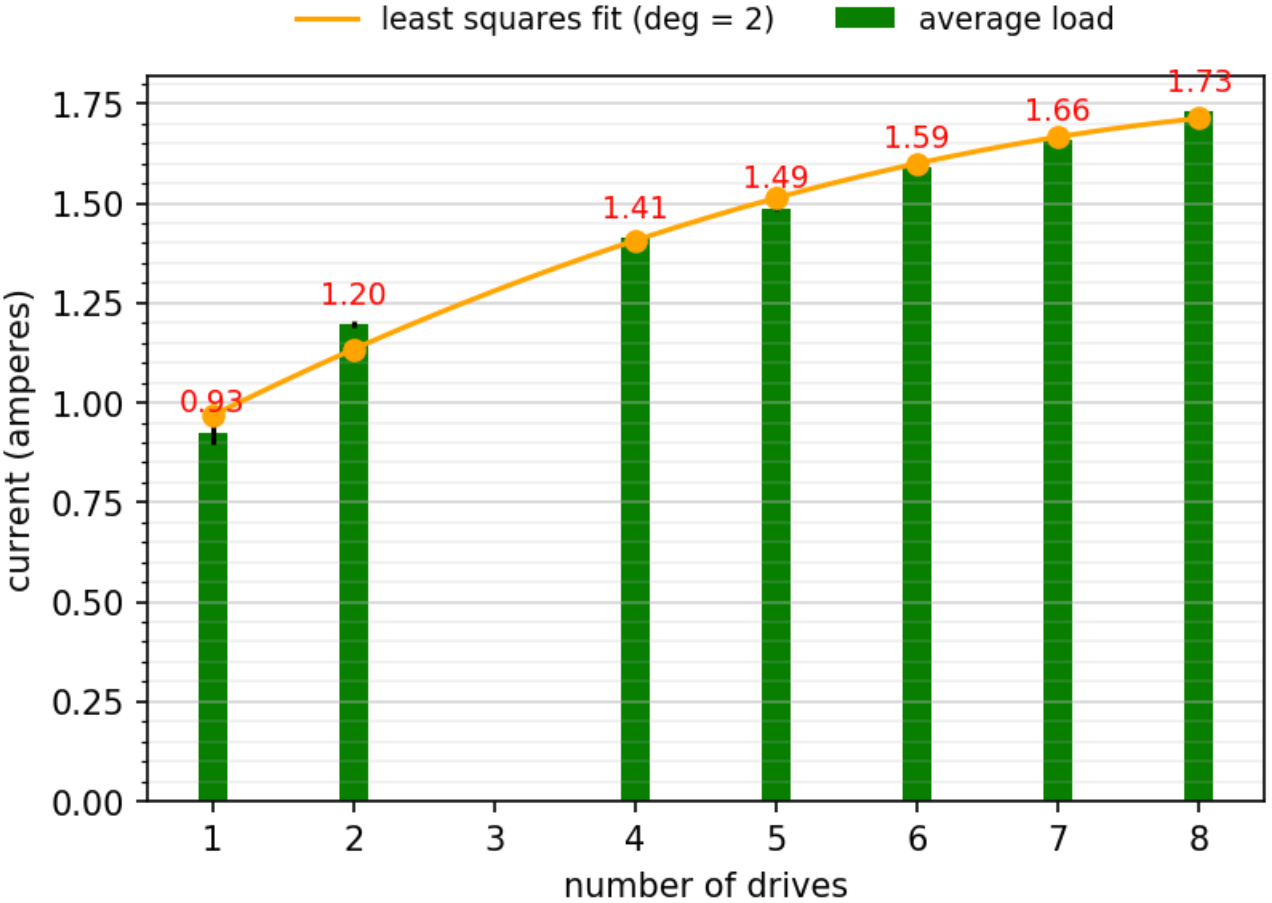}%
    \label{fig:integrated_test_power}%
  }%
  \caption{Integrated Test: Performance of the Host Platform with Different Number of Storage Devices}
  \label{fig:integrated_test}
\end{figure}

\begin{figure}[h!]
  \centerline{\includegraphics[scale=0.18]{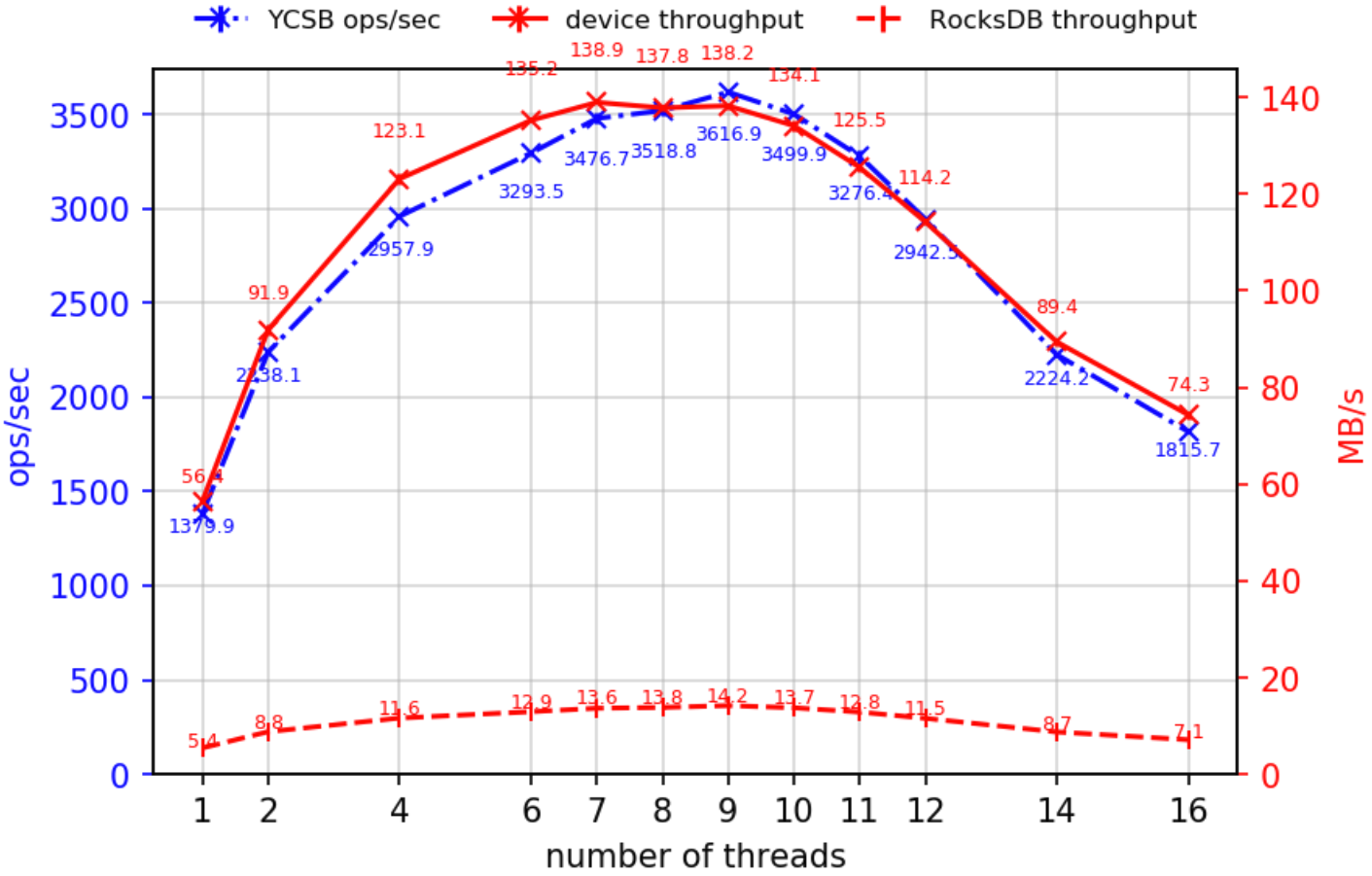}}
  \caption{Integrated Test: Evaluation for the Embedded Platform}
  \label{fig:integrated_test_rockpro64}
\end{figure}

The first tests were integrated tests. Both YCSB and RocksDB ran within platforms and access data from direct-attached storage devices. The purpose of these tests was to reveal the benefit of leveraging cost-effective hardware to provide the function of a key-value data store. The first step was to measure the value of an MBWU. This measurement process went from using one YCSB thread to generate the defined key-value workload to using 32 YCSB threads for concurrent request generation. The reason we stopped at 32 threads relates to the use of SATA SSDs in our experiment. SATA interface provides a single command queue for a depth of up to 32. This feature suggests there is no need to use more than 32 threads to generate requests if the request generation is not slower than the request consumed by the underlying storage device. Figure \ref{fig:data_traffic_amp} shows that the YCSB throughput was mostly stable with more than 20 threads. Considering the amplification factor between the traffic generated by YCSB and the traffic to the underlying storage device, we thought there was no need to test with more than 32 threads. However, if the evaluation regards faster storage devices such as NVMe SSDs, we may need to increase the thread number corresponding to the capability of the storage interface to measure an MBWU. In our results, we saw the YCSB reaches peak throughput with 32 threads on the host platform. We ensured this throughput was the MBWU by carefully examining the resource utilization report generated by our evaluation prototype. Once a single MBWU is measured, we can measure the MBWUs for the two platforms. Our host platform can host up to eight SSDs because it has limited hardware connectors. The workload performance with eight SSDs, as expected, was neither limited by the CPU nor the memory. As discussed previously, the real estate issue is one type of system bottlenecks. This type of bottleneck causes the other system resources to be underutilized; thus, it is conceivable that it increased the values of all three metrics (\$/MBWU, kW/MBWU and $m^{3}$/MBWU) for this platform. Under this restriction, our host platform can generate six MBWUs. Figure \ref{fig:integrated_test} shows the evaluation results of the host platform. \addition{ We skipped some data points in this and some of the following figures as we believed that those values could not be the peak performance numbers according to the trends.} We applied the same measurement methodology to the embedded platform, and the results are shown in Figure \ref{fig:integrated_test_rockpro64}. Limited by CPU performance of this platform, it can only generate 0.5 MBWU with a single SSD. After platform MBWUs are measured, we can compare these platforms using any of the three MBWU-based metrics. We saw that the embedded platform reduced the \$/MBWU by 64\% compared to processing the same key-value workload on the host platform. At the same time, this platform reduced the kW/MBWU by 39.6\% as well. These optimistic results show that it is worth offloading the key-value data management function to the embedded platform due to the significant saving from the hardware. Specifically, compared with the expensive and powerful resources used in the host platform, the cost reduction of the less powerful resources used in the embedded platform is greater than the performance reduction of these resources. In other words, it simply emphasizes the fact that improving the system performance through scaling out is much more cost-effective than through scaling up.

\begin{figure}[h!]
  \centering
  \subfigure[Aggregated Throughput]{%
    \includegraphics[width=0.49\textwidth]{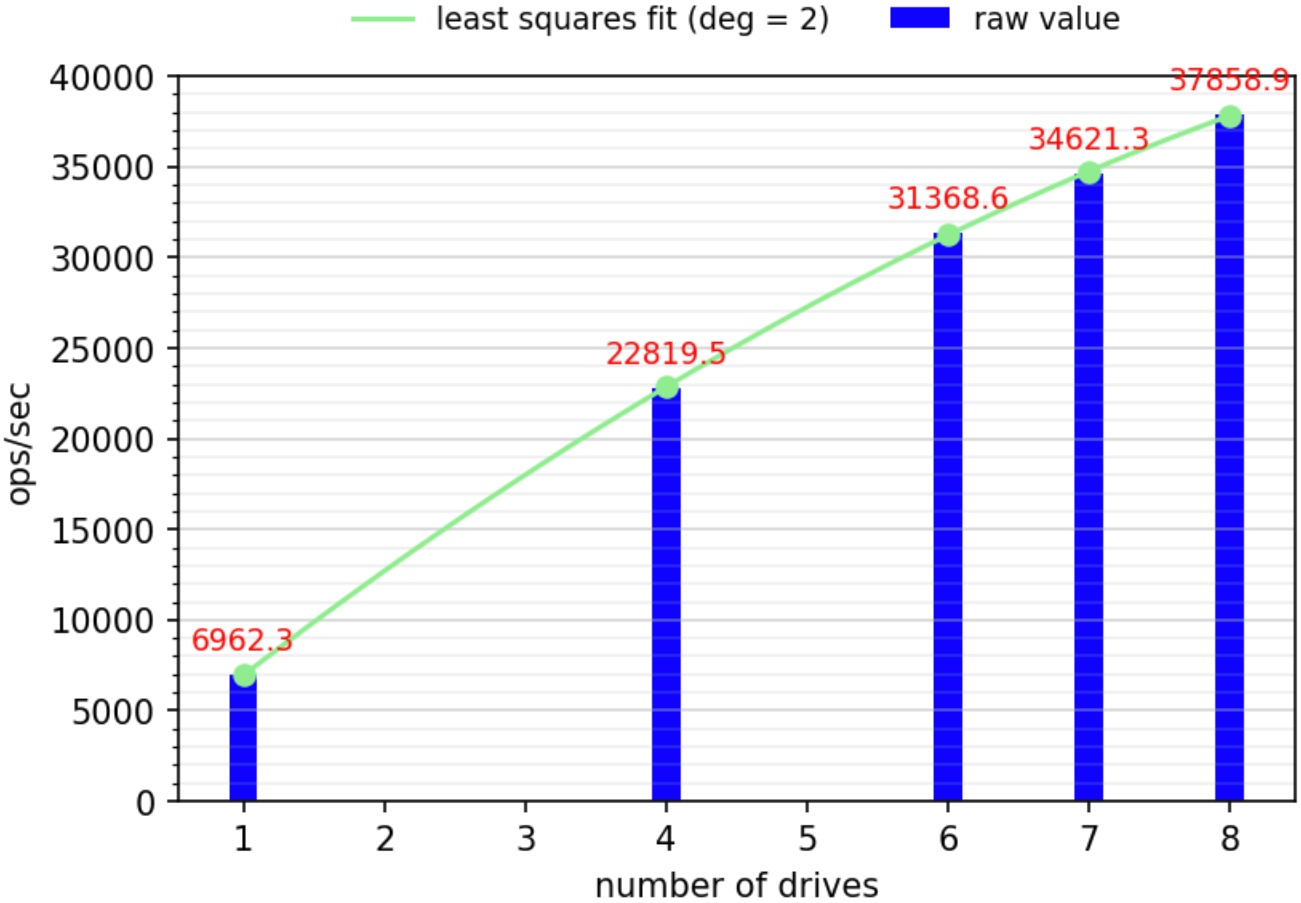}%
    \label{fig:network_test_throughput}%
    }\hspace{0.2cm}
    \subfigure[Platform Power Consumption]{%
    \includegraphics[width=0.49\textwidth]{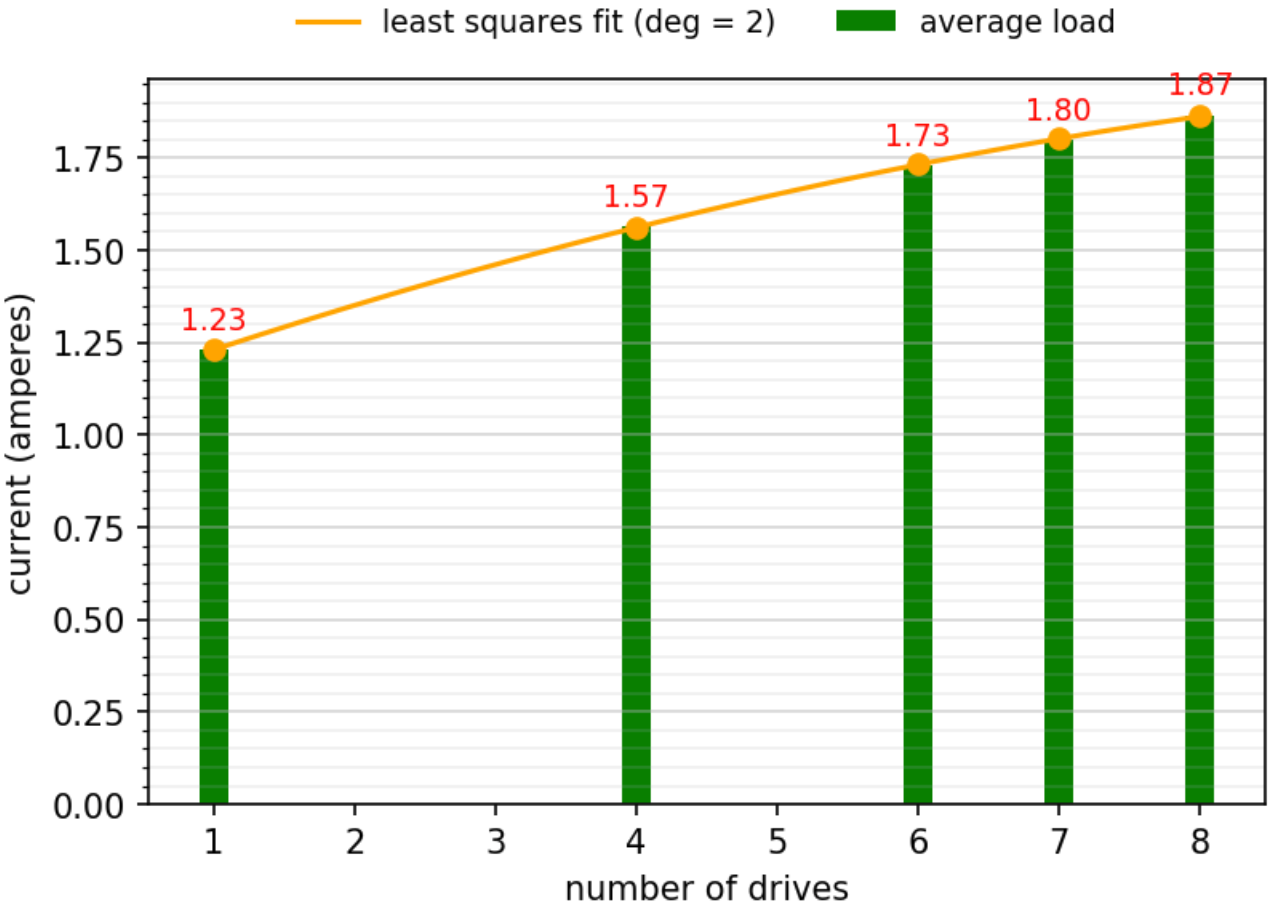}%
    \label{fig:network_test_power}%
  }%
  \caption{Network Test: Performance of the Host Platform with Different Number of Storage Devices}
  \label{fig:network_test}
\end{figure}

\begin{figure}[h!]
  \centerline{\includegraphics[scale=0.18]{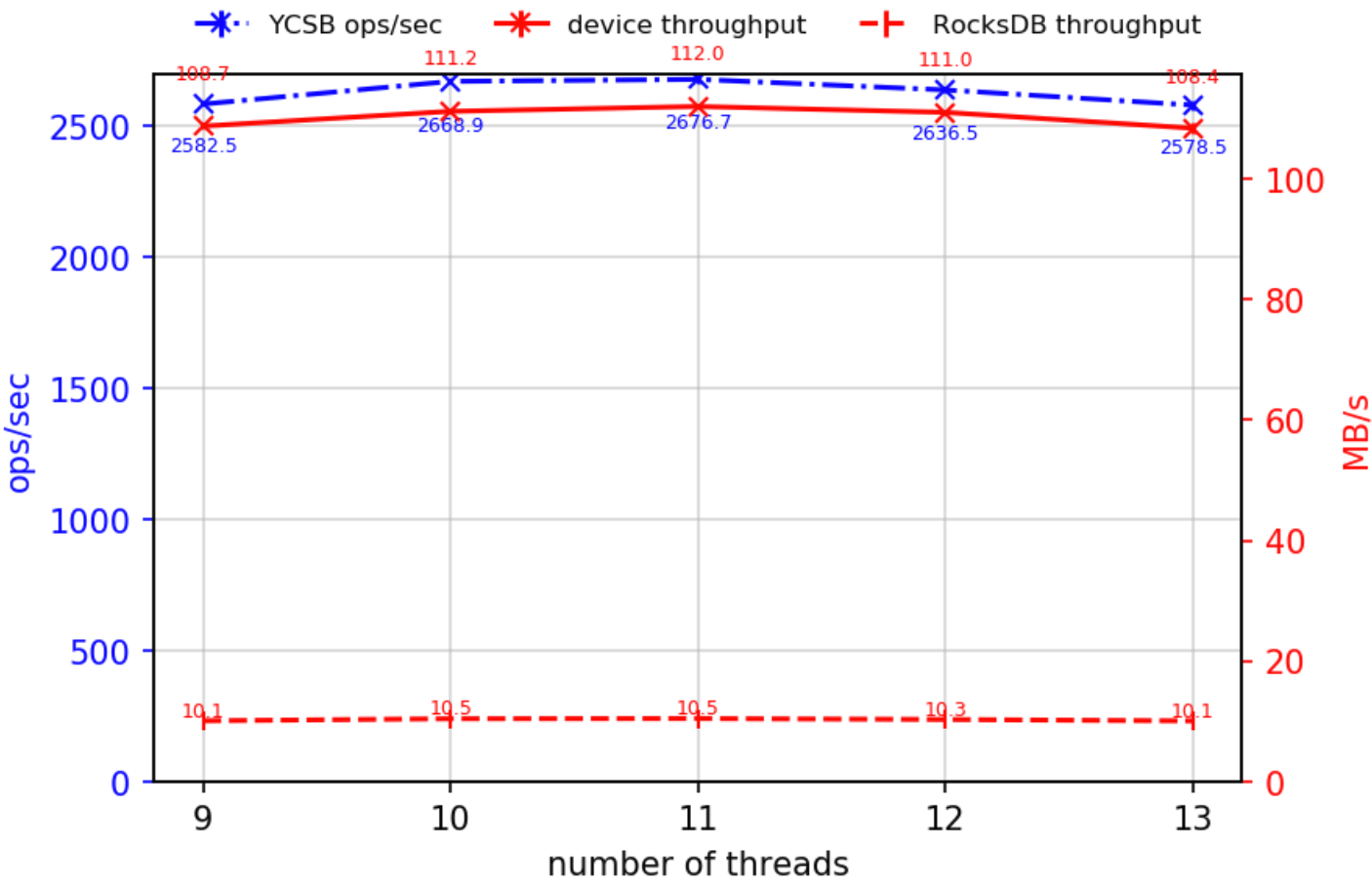}}
  \caption{Network Test: Evaluation for the Embedded Platform}
  \label{fig:network_test_rockpro64}
\end{figure}

In network tests, YCSB sent key-value requests through network as opposed to through local bus in the integrated tests. The introduction of network traffic may have different impacts on different platforms depending on the availability of computing resources and the amount of network traffic. For our host platform, since its throughput performance was not CPU or memory bound in the integrated tests, adding the overhead to process network packets has lower performance impact than the embedded platform where its CPU was already the performance bottleneck for the defined key-value workload. Therefore, the purpose of the network tests was to evaluate how the introduction of this front-end network affected the benefit results we obtained from the integrated tests. Figure \ref{fig:network_test} and \ref{fig:network_test_rockpro64} respectively show the results of our host platform and embedded platform for these tests. Based on the results, the host platform can generate 5.2 MBWUs, and the embedded platform can generate 0.37 MBWUs. With these numbers, we again compared the two platforms using the \$/MBWU and the kW/MBWU metrics. We saw that the embedded platform saved 57.86\% of \$/MBWU compared to processing the same key-value workload on the host platform. For energy consumption, the embedded platform can reduce the kW/MBWU by 45.9\% as well. The reduction of benefit on \$/MBWU is expected since the performance degradation on the embedded platform is greater than the performance degradation on the host platform. Similarly, the percentage of energy saving was increased because the host platform utilized additional system resources for network traffic processing, which raised its total power consumption. For the embedded platform, it had already enabled all system resources to process the workload. In other words, the embedded platform was already under the peak power consumption no matter whether it was required to deal with network traffic.

\begin{figure}[h!]
  \centering
  \subfigure[Aggregated Throughput]{%
    \includegraphics[width=0.5\textwidth]{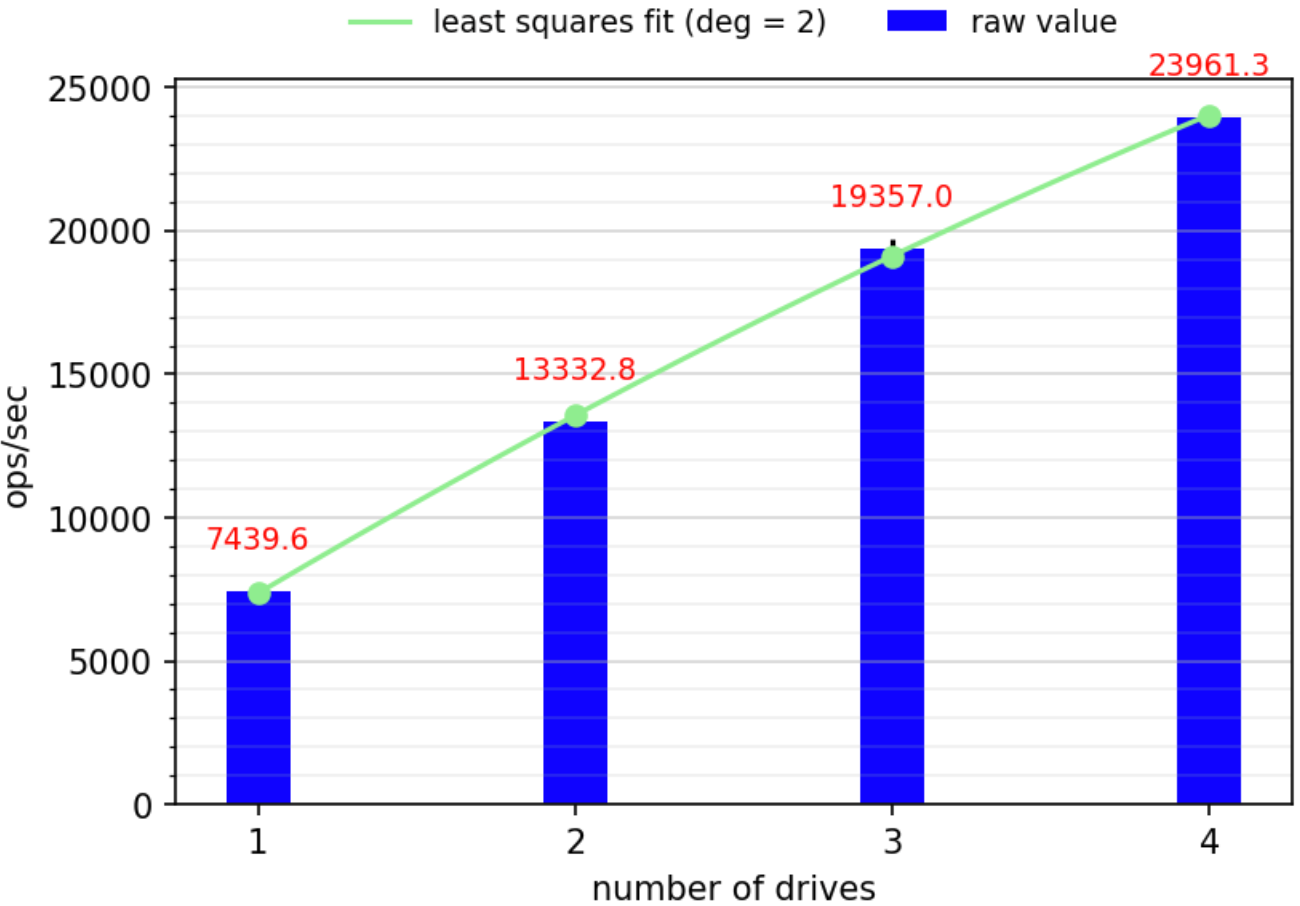}%
    \label{fig:disaggregated_test_throughput}%
    }\hspace{0.2cm}
    \subfigure[Platform Power Consumption]{%
    \includegraphics[width=0.48\textwidth]{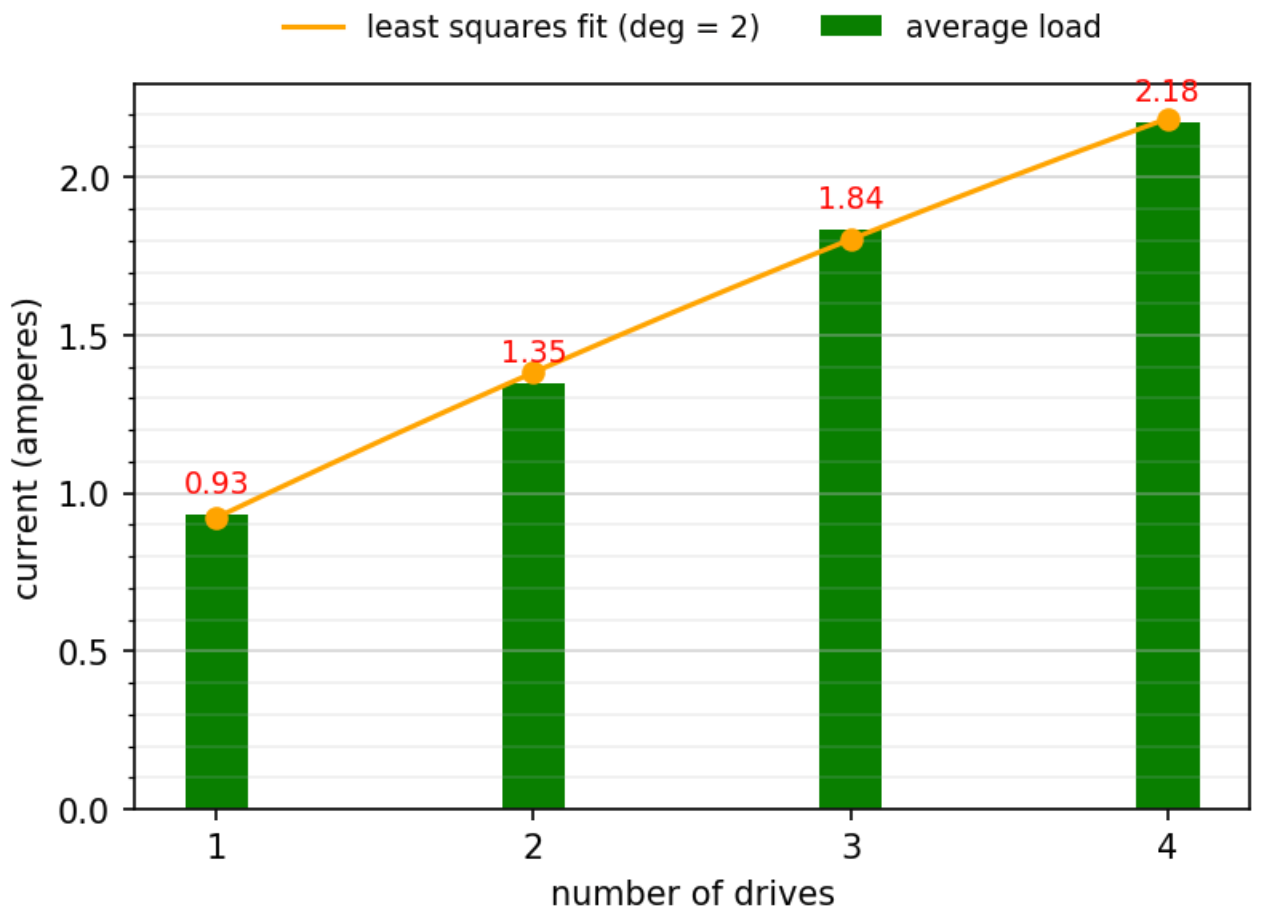}%
    \label{fig:disaggregated_test_power}%
  }%
  \caption{Disaggregated Test: Performance of the Host Platform with Different Number of Storage Devices}
  \label{fig:disaggregated_test}
\end{figure}

Storage disaggregation is known for simplifying and reducing the cost of storage management. It requires additional expense on network infrastructure as the data lives remotely. The faster the storage devices, the higher network bandwidth is required. Hiding the data management traffic within storage devices is especially attractive in this context due to the high amplification factor for data access---6x amplification means more than 5x extra expense on the network to support the bandwidth that is not directly relevant to user applications. In the case of key-value data management function, data amplification comes from data compaction and compression that frequently happen behind the scenes of client applications. In the last test setup, we simulated an environment with disaggregated storage devices to evaluate how much we can save from removing the back-end network requirement for data management traffic. The host and storage devices are connected using iSCSI. Our host platform exactly captured the cost overhead resulting from the data amplification. On the one hand, the built-in network interface card (NIC) in our host platform was unable to support the high bandwidth requirement of the back-end network; we had to install a capable NIC on it, which increased the cost of this platform. On the other hand, the new NIC occupied a PCIe slot causing the reduction of the number of available connectors for storage devices to 4. This reduction exacerbated the unbalance of system resource utilization on this platform and resulted in a lower MBWUs number that the platform can generate, thus decreasing the platform efficiency represented by the three MBWU-based metrics. It is worth mentioning that keeping the system resource utilization in balance for transitional platforms is practically untraceable. In the HPC environment, the ratios between different components in a traditional platform (e.g., the ratio between the number of CPU cores and the number of NICs, and the ratio between the size of memory and the number of storage devices) were designed at the time of purchase according to the requirements of expected workloads. However, the change of workloads is difficult to predict; should that change, the system resource utilization could easily become unbalanced. Figure \ref{fig:disaggregated_test} shows the performance results of our host. In disaggregated tests, the host platform could generate only 3.28 MBWUs. The number of MBWUs of the embedded platform is the same as in the network tests since its setup is the same. By putting all these numbers together, the embedded platform can save 73.4\% of \$/MBWU and 70.7\% of kW/MBWU if we choose not to use the host platform to process the target key-value workload.

\section{Related Work}

Choi et al. ~\cite{choi2015energy} evaluated the energy efficiency of scale-in clusters that support in-storage processing using computation and data-movement energy models. Do et al. ~\cite{do2013query} suggested offloading selected query processing components to smart SSDs. The comparisons were conducted based on raw performance metrics such as elapsed time and energy in Joules, and did not involve any cost comparison. Floem ~\cite{phothilimthana2018floem} is a programming system that aims to accelerate NIC applications development by providing abstractions to ease NIC-offloading design. Biscuit ~\cite{gu2016biscuit} is a near-data processing framework. It allows developers to write data-intensive programs to be offloaded onto storage devices. Both Floem and Biscuit are similar to our evaluation prototype in that they provide a way to trial and error instead of modeling, which is helpful given the complexity of real-world hardware environment. Our MBWU-based measurement methodology differs from all the previous research in that it focuses on quantifying the benefit of offloading alternatives.

\modification{

\section{Conclusion}

Host platforms and embedded platforms differ greatly in resource allocations and placements, causing the cost per performance to be significantly different under the same workload. To quantify the benefit of offloading a given data access function to an embedded platform, we proposed a novel MBWU-based measurement methodology. The core of this methodology is to construct an MBWU as a workload-dependent and media-based reference point and use the MBWU to normalize the performance of different platforms such that the performance values of these platforms are directly comparable. It is the direct comparability that enables us to perform apple-to-apple efficiency comparisons for different platforms. Our evaluation prototype releases the power of this methodology and automates the evaluation process for quantifying the benefit of offloading the key-value data management function under a customized workload. Our next step is to evaluate the benefit of offloading other types of data access functions, such as data decryption/encryption functions, database select/project functions, and other data management functions. We believe this measurement methodology will be a useful tool as it fills the need for benefit quantification in current in-storage computing development.

}

\bibliographystyle{splncs04}
\bibliography{refs}
\end{document}